\documentstyle[12pt]{article}

\newcommand{\be}{\begin{equation}}
\newcommand{\ee}{\end{equation}}
\begin{document}

\begin{center}
{\Large {\bf \ LOCAL REALISTIC THEORY FOR PDC EXPERIMENTS BASED ON THE WIGNER
FORMALISM}}

\vspace{1cm}

{\bf Alberto Casado, Ram\'on Risco-Delgado and Emilio Santos}$^a$.

Escuela Superior de Ingenieros, Universidad de Sevilla,

41092 Sevilla, Spain.

$^a$Departamento de F\'\i sica Moderna, Universidad de Cantabria,

39005 Santander, Spain.

\vspace{1cm}

{\it Presented at the 3rd Workshop on Mysteries, Puzzles and Paradoxes in Quantum mechanics, Gargnano, Italy, September 17-23, 2000}

\vspace{1cm}

{\bf Abstract}
\end{center}

In this article we present a local hidden variables model for all
experiments involving photon pairs produced in parametric down conversion,
based on the Wigner representation of the radiation field. A modification of
the standard quantum theory of detection is made in order to give a local
realist explanation of the counting rates in photodetectors. This model
involves the existence of a real zeropoint field, such that the vacumm level
of radiation lies below the threshold of the detectors.

\vspace{1cm}


\newpage

\section{Introduction}

Following Bell, a local hidden variables model (LHV) exists for an
Einstein-Podilsky-Rosen (EPR) experiment if it is possible to write the
single and joint detection probabilities in the form \cite{Bell}

\begin{equation}
p_{i}=\int \rho (\lambda )P_{i}(\lambda ,\phi _{i})d\lambda \ \quad ;\quad
p_{12}=\int \rho (\lambda )P_{1}(\lambda ,\phi _{1})P_{2}(\lambda ,\phi
_{2})d\lambda ,  \label{1}
\end{equation}
$\lambda $ being the hidden variables with a probability distribution $\rho
(\lambda ).$ $P_{i}(\lambda ,\phi _{i})$ $(i=1,2),$ are some functions
fulfilling the conditions $0\leq P_{i}(\lambda ,\phi _{i})\leq 1$. $\phi
_{1} $ and $\phi _{2}$ represent controllable parameters of the experimental
setup.

Many experiments have been performed in order to test local realism vs.
quantum mechanics via Bell's inequalities, but none of these experiments
have violated Bell's genuine inequalities, which are based upon the
assumptions of realism and locality alone. All inequalities tested up to now
involve additional assumptions, like {\em no-enhancement}. In the last few
years \cite{horne} experiments using photon pairs produced in parametric
down conversion (PDC) have become very popular. LHV models for EPR
experiments with PDC have appeared in the literature \cite{santospdc}, the
aim of which has been to show that local realism had not been ruled out by
these experiments, and to stress the relevance of the so called
``loopholes'', in particular the one due to the low efficiency of optical
photon counters. However, those models were mathematical constructs without
physical content; they have no predictive power.

The purpose of this paper is to exhibit a physical model which is based on
the Wigner formulation of PDC. Our previous work \cite{pdc1} has shown that
the Wigner function of the electromagnetic field is positive for all
performed experiments, and hence it provides a realistic description for the 
{\em production} and {\em propagation} of the radiation in these
experiments. In particular, the Wigner function of the vacuum is the gaussian

\begin{equation}
W(\{\alpha _{{\bf k}},\alpha _{{\bf k}}^{*}\})=\prod_{{\bf k}}\frac{2}{\pi }%
e^{-2|\alpha _{{\bf k}}|^{2}},  \label{w}
\label{nueva}
\end{equation}
which in our local realistic model is interpreted as the probability
distribution of a random radiation filling the whole space (the zeropoint
field, ZPF). Thus the amplitudes \{$\alpha _{{\bf k}}\}$ play the role of
the hidden variables $\lambda $ and W the role of the function $\rho
(\lambda )$ in Eq.(\ref{1}). The nonlinear crystal, acted by a laser beam,
gives an outgoing PDC\ field

\[
E^{(+)}=\sum_{{\bf k}}\left( \frac{\hbar \omega }{\epsilon _{0}L_{0}^{3}}%
\right) ^{1/2}[\alpha _{{\bf k}}e^{-i{\bf k}\cdot {\bf r}+i\omega t}+g\alpha
_{{\bf k}}^{*}e^{-i({\bf k}_{0}-{\bf k)}\cdot {\bf r}+i(\omega
_{0}-\omega )t}+\frac{1}{2}g^{2}\alpha _{{\bf k}}e^{-i{\bf k}\cdot {\bf r}%
+i\omega t}], 
\]
where the first term is the ZPF that crosses the crystal without any
modification, and the other terms are produced via the non-linear coupling
between the laser and the ZPF ($g$ is the coupling parameter).

The PDC beam propagates through a number of devices (lenses, beam splitters,
etc.) characteristic of every experiment and finally some intensity I$_{j}$
arrives at every detector j. The intensity I$_{j}$ is a function bilinear of
the amplitudes $\alpha _{{\bf k}}$ and $\alpha _{{\bf k}}^{*}$ of Eq.(\ref{1}%
). These functions I$_{j}$ are quite involved and have been studied in
detail in Refs.\cite{pdc1} for many experiments. The single and joint
detection probabilities are given by

\begin{equation}
p_{j}=\int W(\{\alpha _{{\bf k}}\},\{\alpha _{{\bf k}}^{*}\})Q_{j}(\{\alpha
_{{\bf k}}\},\{\alpha _{{\bf k}}^{*}\},\phi )d^{N}\alpha _{{\bf k}%
}d^{N}\alpha _{{\bf k}}^{*}\;,  \label{641}
\end{equation}
\begin{equation}
p_{12}=\int W(\{\alpha _{{\bf k}}\},\{\alpha _{{\bf k}}^{*}\})Q_{1}(\{\alpha
_{{\bf k}}\},\{\alpha _{{\bf k}}^{*}\},\phi _{1})Q_{2}(\{\alpha _{{\bf k}%
}\},\{\alpha _{{\bf k}}^{*}\},\phi _{2})d^{N}\alpha _{{\bf k}}d^{N}\alpha _{%
{\bf k}}^{*},\;  \label{64}
\end{equation}
where 
\begin{equation}
Q=\frac{\eta }{h\nu }\int dt\int d^{2}r\left[ I(\{\alpha _{{\bf k}%
}\},\{\alpha _{{\bf k}}^{*}\},\phi ,{\bf r},t)-I_{0}\right] ,  \label{itilde}
\label{masnueva}
\end{equation}
$I=E^{+}E^{-}$ being the intensity of the field arriving at the
corresponding detector and $I_{0}$ is a constant which in our model is
interpreted as the mean intensity of the the ZPF. The integration is carried
over the time window and the surface aperture of the detector. We have
divided \ref{masnueva} by the typical energy of one photon, so that $Q$ becomes dimensionless, $\eta $ being the quantum efficiency of the detector.

The relevant question is whether (\ref{641}) and (\ref{64}) may be
considered particular cases of (\ref{1}) with Q$_{j}$ playing the role of P$%
_{j}$. The answer is not affirmative because we cannot guarantee that 
$0\leq Q \leq 1$. Consequently we conclude that it is not
possible to interpret directly the Wigner-function formalism as an LHV model
for the PDC experiments.

\section{The Detection Model}

We shall devote the rest of the article to the description of the model of a detector that works in a strictly local way and provides the LHV model we are searching for. We will first show the basic points of our model:

\begin{enumerate}
\item  The detector consists of a set of individual photodetector elements, $%
D_{l}$, each characterized by a frequency $\omega _{l}$, and a wave vector $%
{\bf k}_{l}$ ($\omega _{l}=|{\bf k}_{l}|/c$), to which $D_{l}$ responds. We
shall consider the direction of ${\bf k}_{l}$ to be normal to the surface of
the detector, which is taken as a cylinder of area $\pi R^{2}$ and length $L$.

\item  The relevant quantity for the detection is a filtered field
corresponding to a detector element $D_{j}$: 
\begin{equation}
\overline{E}_{l}^{(+)}=\frac{1}{\pi R^{2}LT}\int_{V}dV\int_{0}^{T}E^{(+)}(%
{\bf r},t)\mbox{e}^{-i{\bf k}_{l}\cdot {\bf r}+i\omega _{l}t}dt.  \label{2'}
\end{equation}
Hence it follows that the photodectector element $D_{l}$ is sensitive to
radiation with frequencies in the interval $(\omega _{l}-\frac{\Delta \omega 
}{2},\omega _{l}+\frac{\Delta \omega }{2})$ with $\Delta \omega \approx 
\frac{2\pi }{T},$ $T$ being the detection time window. If we assume that the
incoming light beam has frequencies in the interval ($\omega _{\min },\omega
_{\max })$ with an average frequency $\overline{\omega }=(\omega
_{max}+\omega _{min})/2,$ and $\tau $ is the coherence time of the beam, we
shall have $\delta \omega \equiv \omega _{\max }-\omega _{\min }\approx 2\pi
/\tau ,$ so that the minimum number of detecting elements is $N\approx
\delta \omega /\Delta \omega \approx T/\tau .$ By putting typical values, $%
T=10^{-8}\,s$ and $\tau =10^{-12}\,s$, we have the condition $N>10^{4}$.

\item  We now define the {\em effective intensity} obtained from the
filtered fields in the form

\begin{equation}
\overline{I}=c\epsilon _{0}\sum_{l=1}^{N}\overline{E}_{l}^{(+)}\overline{E}%
_{l}^{(-)}.  \label{imed}
\end{equation}
After that we replace the standard quantum Eq.(\ref{itilde}) by the
expression

\begin{equation}
Q(\overline{I})=(1-\mbox{e}^{-\zeta (\overline{I}-\overline{I}_{0})})\Theta (%
\overline{I}-I_{m})\quad ;\quad \zeta =\eta (h\nu )^{-1},  \label{nos}
\end{equation}

which completes the definition of our model. $\overline{I}_{0}$ is the
average of $\overline{I}$ for the ZPF. $I_{m}$ is some threshold intensity
fulfilling the condition $I_{m}>\overline{I}_{0}$, and $\Theta (x)$ is the
Heaviside function.

\item  Now we rewrite the detection probabilities $\left( \ref{641}\right) $
and $\left( \ref{64}\right) $ in the following equivalent form 
\begin{equation}
p=\int \rho (\overline{I})Q(\overline{I})d\overline{I}\quad ;\quad
p_{12}=\int \rho _{12}(\overline{I}_{1},\overline{I}_{2})Q_{1}(\overline{I}%
_{1})Q_{2}(\overline{I}_{2})d\overline{I}_{1}d\overline{I}_{2.}  \label{noso}
\end{equation}
\end{enumerate}

The probability distribution of the {\em effective intensity} is a
gaussian, which is determined by the mean and the standard deviation (for
details see \cite{detec}). For instance, the probability distribution for
the zeropoint field is

\begin{equation}
\rho _{0}(\overline{I})=\frac{1}{\sqrt{2\pi }\sigma _{0}}\mbox{e}^{(%
\overline{I}-\overline{I}_{0})^{2}/2\sigma _{0}^{2}};\quad \overline{I}_{0}=%
\frac{\overline{\omega }\delta \omega }{8\pi cL}\ \ ;\ \sigma _{0}=\overline{%
I}_{0}\sqrt{\frac{\tau }{T}}.  \label{vacio}
\end{equation}

The probability distribution of the effective intensity when there is a PDC
signal present, $\rho (\overline{I}),$ is similar to that of the ZPF, but
the following remarks are in order: $(a)$ We shall define the signal mean
effective intensity by $\overline{I}_{s}=\overline{\langle I}\rangle -%
\overline{I}_{0}.$ Here the word {\it mean} refers to both, time average
over the detection window and ensemble average with the distribution
function Eq.(\ref{nueva}). If we assume that the signal enters parallel to the axis of the detector, it can be shown that the effective intensity, $%
\overline{I}_{s},$ is equal to the actual intensity, i.e. $\overline{I}%
_{s}\approx I_{s}.$ In practical situations the relation $\overline{I}%
_{s}\ll \overline{I}_{0}$ is fulfilled. $(b)\ $The relation between the mean
and the standard deviation is completely analogous to that of the ZPF alone
(see Eq. (\ref{vacio})), and the distribution of the effective intensity $%
\overline{I}$ may be written

\begin{equation}
\rho (\overline{I})=\frac{\sqrt{T/\tau }}{\sqrt{2\pi }\langle \overline{I}%
_{0}\rangle }\mbox{e}^{-\frac{(\overline{I}-\overline{I}_{0}-\overline{I}%
_{s})^{2}T}{2\overline{I}_{0}{}^{2}\tau }}.  \label{rodei}
\end{equation}

On the other hand, $\rho (\overline{I}_1,\overline{I}_2)$ is a doble
gaussian function which is defined by the mean values of its marginals,
their standard deviations and the correlation function $\langle (\overline{I}%
_1-\overline{I}_{1s}-\overline{I}_{10})(\overline{I}_2-\overline{I}_{2s}-%
\overline{I}_{20})\rangle $.

\section{The Detection Probabilities}

In this section we shall compare the predicted detection probabilities of our model with those of quantum optics. In the case of single detection probability, let us consider the three following possible situations:

{\em i)} $\bar{I}_{s}=0$. In sharp contrast with quantum optics, our model
predicts the existence of some counts in any detector even in the absence of
signal, the probability being very small if $I_{m}-\overline{I}_{0}\gg
\sigma _{0}.$

{\em ii) }$\zeta \bar{I}_{s}\ll \sigma _{0}.$ This should be the normal
situation in experimental practice. In this case we may choose $I_{m}$ so
that $\overline{I}_{0}+\bar{I}_{s}-I_{m}\gg \sigma _{0}$, but $I_{m}>%
\overline{I}_{0}$ in order to preserve the positivity of $Q$ in Eq. (\ref
{nos}). With these two constraints it can be shown that $p\approx \zeta
I_{s},$ a result that agrees with the quantum one if we put $\zeta =\frac{%
\eta _{i}}{h\nu _{i}}$ .

{\em iii) }$\zeta \bar{I}_{s}\gg \sigma _{0}$. In this case the detector
saturates and gives a count in every time window.

Finally, in the case of the joint detection probability the predictions of
our model coincide with the quantum result in the limits $\frac{\eta _{i}}{%
h\nu _{i}}(\overline{I}_{i}-\overline{I}_{i0})\ll 1$, which requires $%
\overline{I}_{is}+\overline{I}_{i0}-\overline{I}_{im}\gg \sigma _{0}$.

\section{Constraints of the Model}

In our model there is a trade-off between the constraint $\overline{I}_{0}+%
\bar{I}_{s}-I_{m}\gg \sigma _{0}$, required for the linearity of the
response at low efficiency, and $I_{m}-\overline{I}_{0}\gg \sigma _{0},$
needed for the smallness of the dark counting probability. The need to
satisfy both conditions implies that $\overline{I}_{s}>>\sigma _{0}.$ This
means that there is a minimal intensity of the signal which may be reliably
detected, a constraint absent in the quantum theory of detection, but
certainly existing in experimental practice.

Let us analyze the consequences of the constraint. In experimental practice
a lens is placed in front of the detector in such a way that the signal
field has spatial coherence on the surface of the lens. The condition for
having spatial coherence is $d\lambda \geq R_{l}R_{C},$ $d$ being the
typical distance between the nonlinear medium (with radius $R_{C}$) and the
detector; $R_{l}$ is the radius of the lens. On the other hand, the
zeropoint field is not modified by the lens, which is evident because of the
fact that energy cannot be extracted from the vacuum. As a consequence, the
intensity of the incident signal is amplified by a factor $b^{2}\equiv \pi
^{2}R_{l}^{4}/\lambda ^{2}f^{2},$ $f$ being the focal distance. On the other
hand, $84\,\%$ ($91\,\%$) of the total intensity is concentrated within the
first (second) ring of the difraction pattern with a radius $R=a\times
(f\lambda /2R_{l})=a\lambda /A_{r}$, where $A_{r}=2R_{l}/f$ is the relative
aperture of the lens and $a=1.22$ ($2.23$) for the first (second) ring .
Consequently, the optimun radius of the detector is given by $R$.

By taking into account the above considerations, and using Eq. (\ref{vacio})
and the condition $d\lambda \geq R_{l}R_{C}$ , the constraint $\overline{I}%
_{s}\gg \sigma _{0}$ gives 
\begin{equation}
{\rm Rate}>>\frac{\eta f^{2}R_{C}^{2}}{2Ld^{2}\lambda \sqrt{\tau T}},
\label{cond7}
\end{equation}
which puts a lower bound on the single rates which may be used in reliable
experiments. This result cannot be derived from $\left( {\rm conventional}%
\right) $ quantum theory. By putting typical parameters, that is $\eta
\approx 0.1,$ $R_{l}$, $L$ and $f$ of the order of fractions of a centimeter, $
10$ nm (which gives a coherence time $\tau $ $\approx 1$ ps) we get a
minimal counting rate of the order of 10$^{5}-10^{6}$ counts per second.
This figure agrees fairly well with the actual experiments and should be considered a success of our model.

\section{Acknowledgement}

We acknowledge financial support by DGICYT Project No. PB-98-0191 (Spain).

\end{document}